\documentclass[runningheads]{llncs}
\usepackage{graphicx}
\usepackage{amsmath,graphicx}
\usepackage{tikz}
\usetikzlibrary{calc}
\usepackage{comment}
\usepackage{svg}
\usepackage{xcolor}
\usepackage{enumitem}
\usepackage{multirow}

\usepackage{xr}
\usepackage{xr-hyper}

\usepackage[tableposition=top]{caption}

\usepackage[hidelinks]{hyperref}

\usepackage{caption}

\begin{document}
\title{Extremely weakly-supervised blood vessel segmentation with physiologically based synthesis and domain adaptation 
}
%
%
\titlerunning{Vessel segmentation with physiological synthesizing and domain adaptation}
%

\author{Peidi Xu\inst{1} 
\and Olga Sosnovtseva\inst{2} 
\and Charlotte Mehlin Sørensen\inst{2}
\and Kenny Erleben\inst{1} \and Sune Darkner\inst{1}
}

\authorrunning{P. Xu et al.}

\institute{Department of Computer Science,  University of Copenhagen, Denmark \email{peidi@di.ku.dk}\\
\and
Department of Biomedical Sciences, University of Copenhagen, Denmark 
}



\maketitle              
%

\begin{abstract}
Accurate analysis and modeling of renal functions require a precise segmentation of the renal blood vessels. Micro-CT scans provide image data at higher resolutions, making deeper vessels near the renal cortex visible. Although deep-learning-based methods have shown state-of-the-art performance in automatic blood vessel segmentations, they require a large amount of labeled training data. However, voxel-wise labeling in micro-CT scans is extremely time-consuming given the huge volume sizes. To mitigate the problem, we simulate synthetic renal vascular trees physiologically while generating corresponding scans of the simulated trees by training a generative model on unlabeled scans. This enables the generative model to learn the mapping implicitly without the need for explicit functions to emulate the image acquisition process. We 
further propose an additional segmentation branch over the generative model trained on the generated scans. We demonstrate that the model can directly segment blood vessels on real scans and validate our method on both 3D micro-CT scans of rat kidneys and a proof-of-concept experiment on 2D retinal images. 
Code and 3D results are available at 
\footnote{\url{https://github.com/miccai2023anony/RenalVesselSeg}}

\end{abstract}
\section{Introduction}
The vasculature in each organ has a characteristic structure tailored to fulfil the particular requirements of the organ. The renal vasculature, which serves as a resource distribution network, plays a significant part in the kidney's physiology and pathophysiology. Not only does it distribute blood and nutrients to  individual nephrons and regulates the filtration of blood in the kidney, but it also acts as a communication network, allowing neighboring nephrons to interact through electrical signals transmitted along the vessels \cite{marsh2019nephron}. Automatic segmentation of renal blood vessels from medical scans is usually the essential first step for developing realistic computer simulations of renal functions.

\noindent\textbf{General deep-learning based segmentation of blood vessels:}
Deep learning models have been widely used for automatic blood vessel segmentations and have shown state-of-the-art performances applied on lungs, liver, and eyes \cite{jia2021learningbasedvessel,cui2019pulmonary,keshwani2020topnet,zhang2021pyramid}. 
However, only a few efforts were made  for renal blood segmentation. Recently, He et al. proposed Dense biased networks \cite{he2020densebias} to segment renal arteries from abdominal CT angiography (CTA) images by fusing multi-receptive fields and multi-resolution features for the adaptation of scale changes. However, the limited resolution of CTA images only allows the models to reach interlobar arteries and enables the estimation of blood-feeding regions, which is useful for laparoscopic partial nephrectomy but not for analyzing realistic renal functions. Therefore, there is a need for imaging with higher resolution, e.g., micro-computed tomography (micro-CT) scans.

\noindent\textbf{Micro-computed tomography and related deep-learning works:}
 Micro-CT shares the same imaging principle as conventional clinical CT, but with a much smaller field of view, such that microscale images of high spatial resolution can be obtained \cite{ritman2011current}. 
Micro-CT scans are commonly used to study various microstructures including blood vessels \cite{andersen_charlotte_2021evaluation}. Few existing research on the auto-segmentation of micro-CT scans focuses on segmenting internal organs of the heart, spinal cord, right and left lung \cite{malimban2022deep}, and blood vessels on colorectal tissue \cite{ohnishi2021three_mCT_tissue} using either nn-UNet \cite{isensee2021nnunet} or variant 3DUNet \cite{unet_origin,cciccek2016_3dUnet_origin}. There is, however, no prior work in segmenting vasculatures in organs like kidneys from micro-CT scans. 
Crucially, most of the above deep learning methods require a large number of label maps to train the segmentation network. Manual labeling of micro-CT scans is extremely time-consuming given the huge volume size. Therefore, in our case, we do not have any clean label maps to train a segmentation model.

\noindent\textbf{Synthetic training data for blood vessel segmentation:}
Transfer learning from artificially generated data is one possible technique to train deep learning models in a data scarcity setting. The process involves pre-training models on synthetic data, which are then fine-tuned on a small set of labeled real data. In medical image segmentation, this strategy has been widely applied to tumor segmentations \cite{lindner2019using}.
Since blood vessels do follow certain physiological and anatomical properties, e.g., Murray's law \cite{murray1926physiological}, this approach has also been applied to train segmentation models for mouse brain vasculature with physiologically synthesized vessels 
\cite{todorov2020machine}. 
However, these works only pre-train the models on synthetic data and still require real labeled data for fine-tuning. Recently, Menten et al. \cite{menten2022physiology} synthesize retinal vasculature and then emulate the corresponding optical coherence tomography angiography (OCTA) images. They show that a UNet trained on these emulated image-label pairs can directly segment real OCTA images. 
However, the way they generate scans from synthesized labels is completely explicit, which includes a series of physics-based image transformation functions that emulate the image acquisition process (e.g., OCTA). These functions clearly require expert knowledge and do not translate to micro-CT settings.

\noindent\textbf{Generative models for domain adaptation:}
In practice, a relatively large number of unlabeled scans are usually available. Thus, a more general way to generate scans without emulating image acquisition explicitly is to utilize these unlabeled scans via a generative model, i.e., domain adaptation. Medical image segmentation with domain adaptation is an active research area, and a popular method is  Generative Adversarial Networks (GANs).
In particular, CycleGAN \cite{zhu2017cyclegan_orig} has been used to perform domain adaptation for medical image segmentation such as liver and tumor segmentation in CT images, where the goal is to train a segmentation model on one domain and then apply it to another domain with missing or scarce annotated data 
\cite{bui20206,gilbert2021generating}
.  
Recently, Chen et al. \cite{chen2021real} segment cerebral vessels in 2D Laser speckle contrast imaging (LSCI) images using public fundus images with segmentation labels as the source domain and a CycleGAN for domain adaptation.  The ability of CycleGAN to perform translation without paired training data makes it a powerful tool for domain adaptation. 

\noindent\textbf{Our contribution:}
We propose a framework with two main components: 1) a physiology-based simulation that synthesizes renal vascular trees and 2) a generative model with an additional segmentation branch that adapts the synthesized vascular trees to real scans while performing segmentation simultaneously. For 1), we extend the work \cite{xu2023hybrid} from physiologically synthesizing renal arterial tree to venous tree. Since a small prebuilt tree needs to be manually provided in the initialization step of the process, we call our method \textit{extremely weakly-supervised}. For 2), we aim to ``emulate'' corresponding scans using CycleGANs. Specifically, we follow the idea in \cite{chen2021real} to train a vessel segmentation network over the output from CycleGAN while extending to 3D on kidney micro-CT images. 
Notably, although 3D CycleGAN has been adopted for segmenting brain tissues and heart chambers \cite{bui20206,gilbert2021generating,zhang2018translating}, no similar work exists on subtle structures like blood vessels in 3D. Moreovoer, these works still require scans or labels from other sources, modalities, or time points as the source domain. Instead, our source domain is purely physiologically synthesized vascular trees. We show that our combined model can directly segment blood vessels in real scans from the target domain and demonstrate the validity of our approach on segmenting vasculatures both on 3D micro-CT scans of rat kidneys and 2D retinal images.

\section{Method}

\subsection{Physiologically-based simulation of renal blood vessels}
\label{section_physio}

\begin{figure}[htb]
\centering
   \includegraphics[width=.99\textwidth]{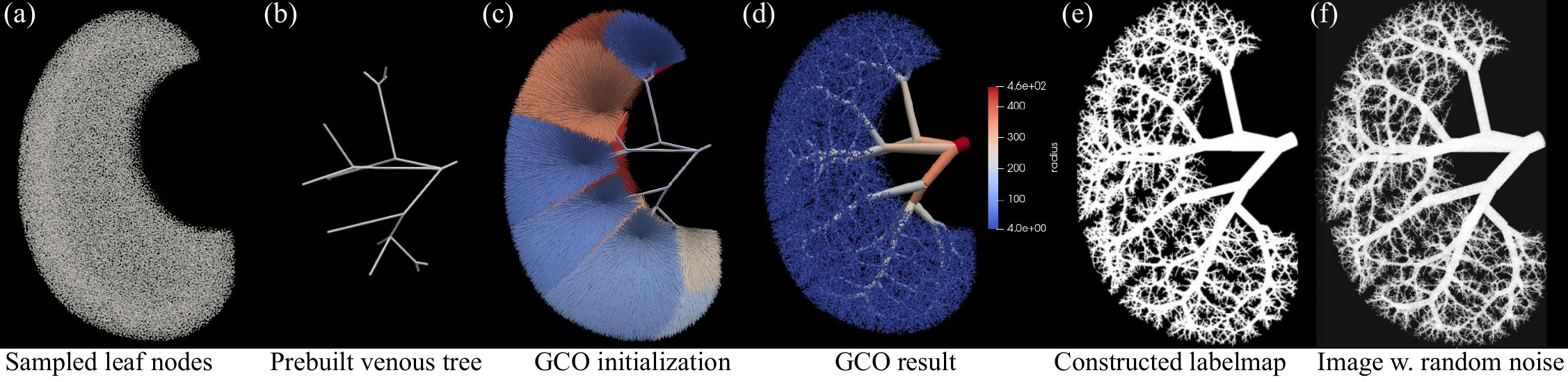}
 \caption{Physiologically based vessel synthesizing pipeline. Details are given in \cite{xu2023hybrid}. The last two subfigures are shown with maximum intensity projection (MIP).}
\label{fig_vessel_construction}
\end{figure}
Constraint Constructive Optimization (CCO) \cite{schreiner2000constrained} and its variant Global Constructive Optimization (GCO) \cite{georg2010global} 
are widely used angiogenesis-based methods that simulate the growth of vascular trees. These methods turn tree growth into an optimization problem based on the biological, physiological, and hemodynamic factors involved in the process.  Here, the vascular tree is modeled by a directed acyclic graph $\mathcal{G} \equiv (\mathcal{V}, \mathcal{E})$ where $\mathcal{V}$ is a set of nodes in the two endpoints of each vessel centerline with its coordinates in Euclidean space as node features, and $\mathcal{E}$ is a set of directed edges representing each vessel segment as a cylindrical tube with its specific radius as edge features. Boundary conditions such as terminal radius and flow distributions are imposed to represent physiologic conditions. The algorithms then find a tree that minimizes the system's overall cost function while fulfilling several constraints. 


Here we follow \cite{xu2023hybrid} which adopts GCO as the backbone model for generating the renal vascular trees with optimal branching structures by performing multi-scale optimizations through iterating several operations such as splitting and relaxation \cite{georg2010global}. However, instead of the arterial tree presented in \cite{xu2023hybrid}, we only focus on the venous tree because it constitutes most of the vessel foreground. In summary, veins follow a similar pattern to arteries but are thicker, which is accomplished by sampling more terminal nodes with large radii (Fig.~\ref{fig_vessel_construction}a). Detailed modifications over boundary conditions from arterial to venous trees are given in supplementary material, including tuning the weighting factors of the loss defined in \cite{xu2023hybrid}. Together with inherent randomnesses in the GCO process itself, the generated tree (Fig.~\ref{fig_vessel_construction}d) will look different each run. This enables a variety of synthesized vascular trees to train the later deep-learning model. 

Note that though a prebuilt tree $\mathcal{G}_0 \equiv (\mathcal{V}_0, \mathcal{E}_0)$ is required to guide the GCO process as noted in \cite{xu2023hybrid}, $\mathcal{G}_0 $ involves less than 20 nodes, as shown in Fig.~\ref{fig_vessel_construction}b, which can be manually selected. This node selection process should take much less time than a voxel-wise manual annotation of the whole blood vessels. Therefore, we call our pipeline \textit{extremely weakly-supervised} with a partially annotated tree structure but without any real segmentation label maps.

\subsection{CycSeg: CycleGAN with additional segmentation branch}
To create a synthetic image dataset, the reconstructed vascular tree structures $\mathcal{G} \equiv (\mathcal{V}, \mathcal{E})$ (Fig.~\ref{fig_vessel_construction}d) are then remapped to a 3D binary label map (Fig.~\ref{fig_vessel_construction}e) by voxelization, the detail of which is given in \cite{xu2023hybrid}.  We then generate the corresponding gray-scale synthetic images (Fig.~\ref{fig_vessel_construction}f) by simply assigning vessel foreground and background with random integers in $[128, 255]$ and $[0, 127]$ respectively. Of course, a segmentation model trained on these image-label pairs (Fig.~\ref{fig_vessel_construction}f \& e) will not work on real scans because of this oversimplified scan construction.

\begin{figure}[htbp]
\centering
          \includegraphics[width=.99\textwidth]{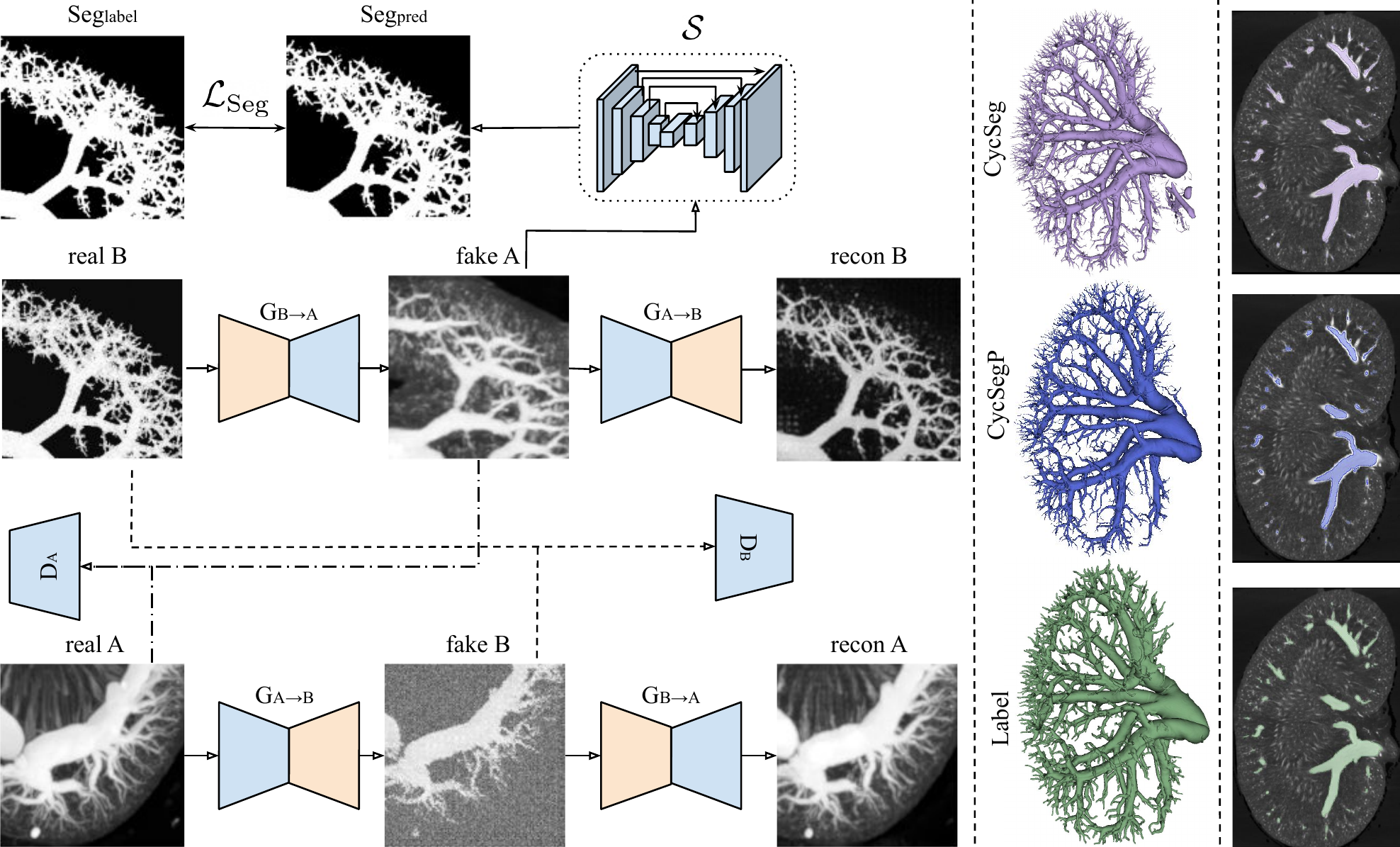}

 \caption{Left: An illustration of CycleGAN with an additional segmentation branch working on adapted (fake) domain A images by $G_{B \rightarrow A}$. All 3D patches are shown with MIP. Computations for $\mathcal{L}_{\text{GAN}}$, $\mathcal{L}_{\text{id}}$, $\mathcal{L}_{\text{cyc}}$ are not shown for simplicity. Middle: An example result in 3D. Right: A sample slice overlaid with segmentation.}
\label{fig_cyclegan_illu}
\end{figure}

In order to adapt corresponding scans that emulate the micro-CT acquisition process out of the label maps from the previous step,  unlabeled real micro-CT scans are utilized to train a generative model. Our backbone model is CycleGAN \cite{zhu2017cyclegan_orig}, which we extend to 3D while integrating an additional segmentation branch with segmenter $S$ using a standard 3D UNet\cite{cciccek2016_3dUnet_origin}, as shown in Fig.~\ref{fig_cyclegan_illu}(Left). We refer to our model as CycSeg later. The only modification we make in 3D is that 32 filters are used in the first layer instead of 64 to ease the computational load. Please refer to the supplementary material for detailed model architectures.

We strictly follow the original design \cite{zhu2017cyclegan_orig} while extending to 3D for the loss functions definition and training procedure. Briefly, the two-way GAN loss $\mathcal{L}_{\mathrm{GAN}}\left(G_{A \rightarrow B}, D_B \right)$ and $\mathcal{L}_{\mathrm{GAN}}\left(G_{B \rightarrow A}, D_A \right)$ is to encourage the generator network to produce samples that are indistinguishable from real samples. Besides, cycle-consistency loss $\mathcal{L}_{\mathrm{cyc}}(G_{A \rightarrow B}, G_{B \rightarrow A})$ is to assure the property ${a \sim p_{\text {data }}(a)}: G_{B \rightarrow A}(G_{A \rightarrow B}(a))=a$ and ${b \sim p_{\text {data }}(b)}: G_{A \rightarrow B}(G_{B \rightarrow A}(b))=b$. In our case, $p_{\text {data }}(a)$ and $p_{\text {data }}(b)$ denotes the real micro-CT scans distribution and the synthetic scans distribution from the physiologically generated trees, respectively. A final identity loss $\mathcal{L}_{\text {id }}(G_{A \rightarrow B}, G_{B \rightarrow A})$ is to stabilize the two generators. Please refer to \cite{zhu2017cyclegan_orig} for the exact definition and computation of the above losses.


In our CycSeg, the additional segmenter ($S$) is trained on fake A (output from $G_{B \rightarrow A}$) with the goal to work on real A, as shown in Fig.~\ref{fig_cyclegan_illu}. This introduces the segmentation loss $\mathcal{L}_{\text{seg}}(S, G_{B \rightarrow A})$. Thus, the final loss $\mathcal{L}_{\text{tot}}$ is defined as
\begin{equation}
\label{eq_tot}
\begin{aligned}
 \mathcal{L}_{\text{tot}} & = \mathcal{L}\left(G_{A \rightarrow B}, G_{B \rightarrow A}, D_A, D_B, S \right) \\
& =\mathcal{L}_{G A N}\left(G_{A \rightarrow B}, D_B \right)  +\mathcal{L}_{G A N}\left(G_{B \rightarrow A}, D_A \right)  \\
& + \lambda_1 \mathcal{L}_{c y c}(G_{A \rightarrow B}, G_{B \rightarrow A})  + \lambda_2 \mathcal{L}_{\text{id}}(G_{A \rightarrow B}, G_{B \rightarrow A})  + \lambda_3 \mathcal{L}_{\text{seg}}(S, G_{B \rightarrow A})
\end{aligned}
\end{equation}
Here  $\mathcal{L}_{\text{seg}}(S, G_{B \rightarrow A})$ is an unweighted combination of a dice loss and standard cross-entropy loss. Specifically, the segmenter $S$ takes the output from the generator $G_{B \rightarrow A}$ as input. Thus, given the physiologically generated label map $g$ and its corresponding synthetic gray-scale image $x \sim p_{\text {data }}(b)$, the segmenter outputs $p=S(G_{B \rightarrow A}(x))$, and $\mathcal{L}_{\text{seg}}(S, G_{B \rightarrow A})$ is defined as
%
%
\begin{equation}
\label{eq_seg}
\mathcal{L}_{\text{seg}}^{}(S, G_{B \rightarrow A}) = \frac{1}{N} \left(-\sum\nolimits_i^N g_i \log p_i + 1-\frac{2 \sum_i^N p_i g_i}{\sum_i^N p_i+\sum_i^N g_i}\right)
\end{equation}
where $N$ is the total number of voxels in each sampled 3D patch.

We follow \cite{zhu2017cyclegan_orig} by setting the weights $\lambda_1=10$ and $\lambda_2=5$ while setting $\lambda_3=3$ experimentally.  Note that although $G_{B \rightarrow A}$ is one of the input models to compute $\mathcal{L}_{\text{seg}}(S, G_{B \rightarrow A})$, all the CycleGAN components including $G_{B \rightarrow A}$ are frozen when training the segmenter $S$ by blocking backpropagation. Moreover, since the generator, discriminator, and segmenter are trained alternately, $\lambda_3$ does not strongly impact the training and only affects early stopping.

During inference, all CycleGAN components are discarded, while real scans (domain A) are directly passed to segmenter $S$ to output segmentation maps. 

\section{Experiments and Results}


\noindent\textbf{Dataset}: The kidney cast was prepared as described in \cite{andersen_charlotte_2021evaluation} in agreement with approved protocols (approval granted from 
the Danish Animal Experiments Inspectorate under the Ministry of Environment and Food, Denmark).
The rat kidneys were \textit {ex vivo} scanned in a ZEISS XRadia 410 Versa $\mu$CT scanner 
with an isotropic voxel size of 22.6 $\mu$m
\cite{andersen_charlotte_2021evaluation}, with a fixed dimension of $1000 \times 1024 \times 1014$. To ease the computational load, scans are auto-cropped to $(692\pm33) \times (542\pm28) \times (917\pm28)$ by intersected bounding cubes from Otsu's thresholding in each dimension \cite{xu2023hybrid}. Here, we use 7 unlabeled scans (domain A) for training and 1 labeled scan for testing, while other 3 unlabeled scans are only validated visually. The synthesized dataset (domain B) has 15 image-label pairs by tuning parameters used in GCO and running multiple times as discussed in Section~\ref{section_physio}.

\noindent\textbf{Pre-processing}: Each generated patch is only preprocessed by simple min-max normalization $X_{\text {scale }}=(x_{i}-x_{\text {min}})/(x_{max}-x_{min})$.

\noindent\textbf{Experimental Setup and Training process}: The network is implemented in PyTorch and trained on NVIDIA A100 with a batch size of 1 and patch size of 208 for 200 epochs. All three components are optimized using the Adam optimizer with the same learning rate of $2\times10^{-4}$ and reduced by 1\% for every epoch.
We apply early stopping if $ \mathcal{L}_{\text{tot}}$ (cf. Eq.\eqref{eq_tot}) of ten consecutive epochs does not decrease. 
Training takes approximately three days to reach convergence, while segmentation during inference takes only around two minutes per scan.


\begin{table}[htb]
\centering
\caption{Segmentation result on private renal and public retina dataset (in $\%$). 
}
\label{tab1}
\begin{tabular}{|cc|l|l|l|l|l|}
\hline
\multicolumn{2}{|c|}{Training data}                                                                                     & Model    & Acc           & DICE          & clDICE    \\ \hline
\multicolumn{1}{|c|}{\multirow{2}{*}{Renal}}  & \multicolumn{1}{|c|}{{\multirow{2}{*}{Synthetic label + Raw scan }}}    & CycSeg   & 99.2$\pm$0.0  & 76.8$\pm$0.3  & 65.0$\pm$0.8    \\
\multicolumn{1}{|c|}{}                        & \multicolumn{1}{|c|}{}                                                  & CycSegP  & 99.2$\pm$0.0  & 77.6$\pm$0.2  & 72.4$\pm$1.0    \\ \hline
\multicolumn{1}{|c|}{\multirow{2}{*}{Retinal}} & CHASE label +  DRIVE image                                             & CycSeg   & 96.1$\pm$0.1  & 74.8$\pm$0.4  & 76.3$\pm$0.1    \\
\multicolumn{1}{|c|}{}                        &  DRIVE label + DRIVE image                                              & UNet2D   & 96.5$\pm$0.1  & 79.6$\pm$0.5  & 78.9$\pm$0.8    \\
\hline
\end{tabular}
\end{table}

\subsection{Results}

As shown in Fig.~\ref{fig_cyclegan_illu}, the CycleGAN successfully adapts realistic noise during micro-CT acquisition to the synthesized images (from real B to fake A), while a UNet trained over the adapted image with the corresponding synthesized label maps can segment real micro-CT scans. Although the ex-vivo scan separates the organ of interest from other parts, the segmentation of venous trees is still challenging due to various noises during the micro-CT acquisition and efficacy of the contrast media. Despite of few false positives, the 3D results in Fig.~\ref{fig_cyclegan_illu} show a smooth and clear venous tree structure, which indicates that the segmenter is trained to recognize veins from noises in real scans by the adapted images from $G_{B\rightarrow A}$.  Since the result should anatomically be a connected tree, we also apply a simple connected component post-processing to the model's output (CycSegP), which successfully removes most of the floating points and gives better results. 

Table~\ref{tab1} shows the numerical evaluation of the segmentation performance on one test image including accuracy, DICE score, and the topology-aware centerline DICE (clDICE) \cite{shit2021cldice}. Although we cannot compare our results with standard segmentation models due to the lack of any labeled training data, we believe that the visual inspection in Fig.~\ref{fig_cyclegan_illu}  and relatively high quantitative results in Table~\ref{tab1} over 3D vessel segmentation task demonstrate the potential of our method in segmenting and building 3D renal models from micro-CT scans.


\subsection{Proof-of-concept on retinal blood vessel segmentation}
We acknowledge that the quantitative analysis 
above is not thorough with only one test data. Thus, we conduct a proof-of-concept on a 2D retinal blood vessel segmentation task. We follow the previous experimental setups but with a patch size of 256 and batch size of 32 due to the ease of computational load in 2D.

\noindent\textbf{Dataset and domain construction}: We adopt the popular DRIVE dataset \cite{staal2004drive_dataset} as the target domain, which includes 40 digital fundus images captured by a Canon CR5 3CCD camera with a resolution of $584\times565$ pixels. However, the 3D vascular structure will not be physiologically correct when projecting to 2D images, and some works argue against compliance with Murray's law in retinal blood vessels \cite{luo2017retinal}. Thus, we do not synthesize retinal blood vessels using a similar physiologically based strategy in kidneys and only focus on the validity of the segmentation power with domain adaptation. Thus, we directly adopt label maps from another dataset (CHASE \cite{fraz2012ensemble}) as the fake source domain. Specifically, we use 20 from 40 images of DRIVE dataset without label maps, together with 28 label maps from CHASE dataset without corresponding images to jointly train the CycSeg. The remaining 20 labeled images from DRIVE are used for testing. Note that we only work on gray-scale images, and label maps from CHASE are resized to $584\times565$ using nearest neighbor interpolation. 

\begin{figure}[htb]
\centering
   \includegraphics[width=.9\textwidth]{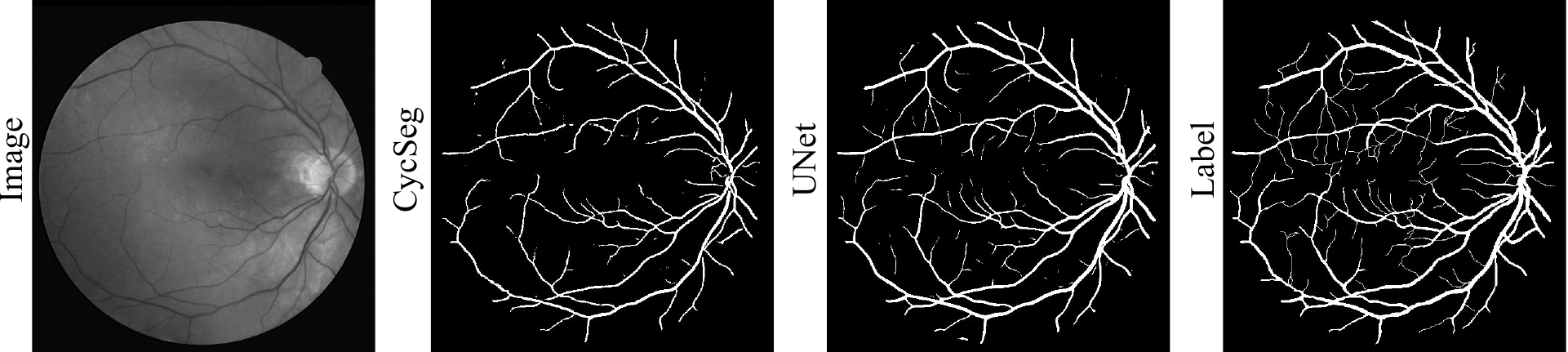}
 \caption{Segmentation results on a DRIVE test image of different strategies.}
\label{fig_2dResult}
\end{figure}

As shown in Fig.~\ref{fig_2dResult}, 
 CycSeg trained on unpaired images is able to segment blood vessels on real images from DRIVE directly. An illustration of retinal domain adaptation in analogue to Fig.~\ref{fig_cyclegan_illu}(Left) is in the supplementary material. As an ablation study, we adopt the same train-test split to directly train a UNet \cite{unet_origin} using real image-label pairs from the DRIVE training set. From Fig.~\ref{fig_2dResult}, both UNet and CycSeg can segment large vessels well while (especially CycSeg) having difficulty detailing small vessels. Numerical results in Table~\ref{tab1} show that the performance of CycSeg trained purely on adapted images is still lower, but the difference is acceptable, as our goal is not to outperform the state-of-the-art but to propose a pipeline that can do segmentation without any labeled images from the target domain. Future work would be to train a UNet using the same generative strategy together with real image-label pairs to see how much information gains it can offer compared with UNet trained only on real image-label pairs.

\section{Conclusion}
We have presented a pipeline that segments blood vessels from real scans without any manually segmented training data. The pipeline first synthesizes label maps of renal vasculatures using a physiologically based model and then generates corresponding scans via a 3D CycleGAN. The CycleGAN emulates image features and artifacts of micro-CT acquisition processes implicitly from unlabeled scans. Simultaneously, an additional segmentation branch on top of CycleGAN trained over the adapted scans enables the segmentation of blood vessels on real scans. 
This removes any need for expert knowledge of scanning settings like proposed in \cite{menten2022physiology}.
We believe our pipeline can crucially reduce annotations needed to segment blood vessels and easily adapt to other organs or modalities and enable computerized diagnosis of vessel-related diseases in clinical practice.

Since segmentation is the final objective, the intermediate image adaptation task is only visually inspected. Future work could include numerical tests like image structure clustering and Amazon
Mechanical Turk perceptual studies \cite{zhu2017cyclegan_orig}.
A modification over \cite{xu2023hybrid} is necessary to model the pair-wise coupling of veins and arteries. This will enable physiologically correct label maps for an Artery/Vein multi-class segmentation task in the future, which is far beyond the scope of this work. Nonetheless, more clean ground truth labels should enable a more thorough validation and benchmark test with standard segmentation models. 





\newpage

%
%
%
%

\bibliographystyle{splncs04}
\bibliography{mybibliography}

\newpage

   \centering \large \textbf{Supplementary Material}

\begin{figure}[htb]
\centering
   \includegraphics[width=.65\textwidth]{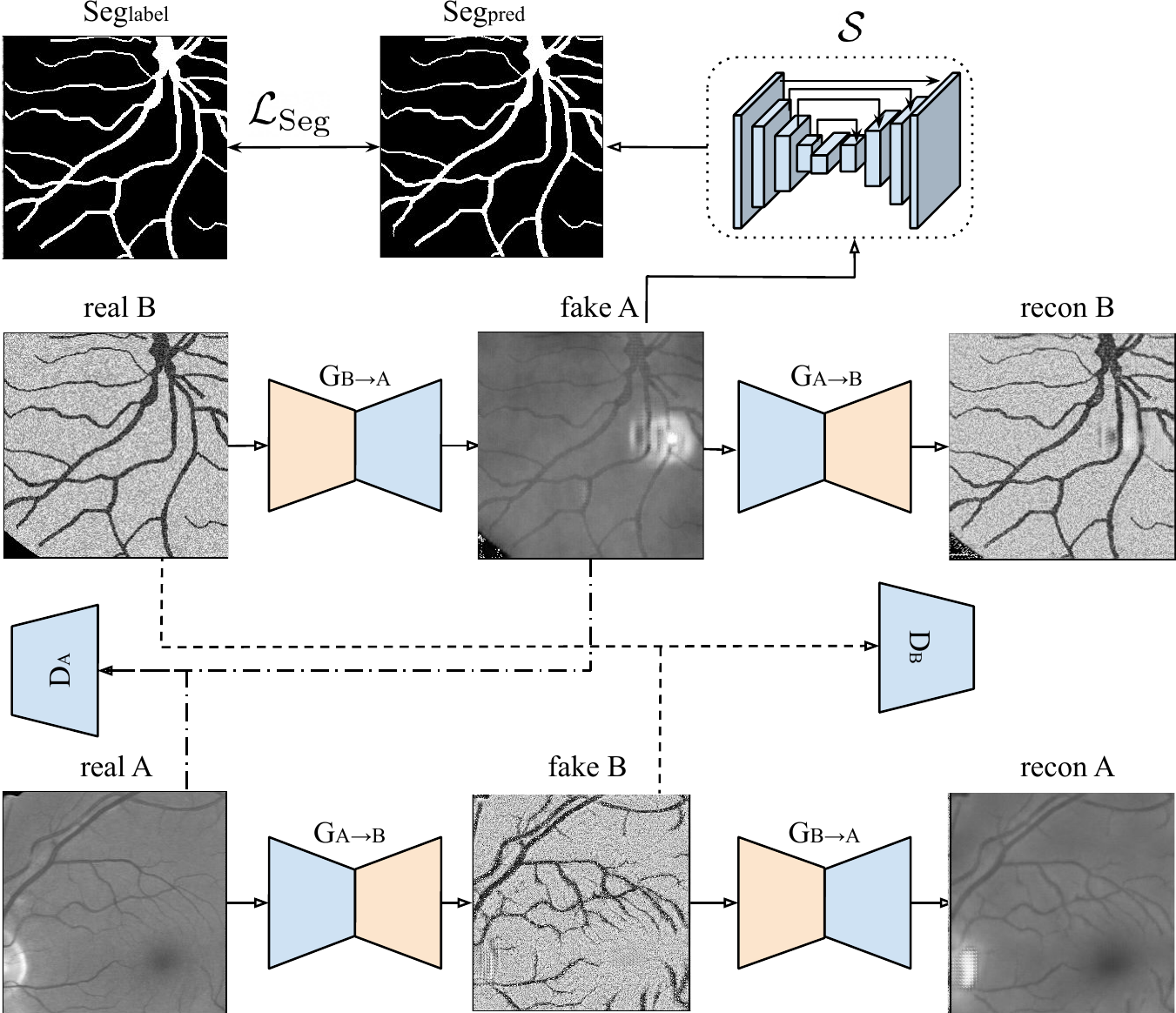}
 \caption{Process over 2D Retinal images. Real Domain B images are simply generated from Domain B labels by assigning vessel foreground and background with random values in $[0, 127]$ and $[128, 255]$ respectively.}
\label{fig_vessel_construction_2d}
\end{figure}


\begin{table}[htb]
\caption{3D Generator structure. pad and out\_pad represent the amount of padding applied in all three dimensions to the input and the output, respectively. Please refer to Github Repo for details such as the usage of InstanceNorm3d and activation functions.}
\centering
\begin{tabular}{|c|c|}
\hline
Layer                                                               & Output feature map size                                   \\ \hline
Input                                                               & $1\times 208\times 208\times 208$                         \\ \hline
$32 \times 7 \times 7 \times 7$ conv, stride 1, pad 3               & $32\times 208\times 208\times 208$                       \\ \hline
$64 \times 3 \times 3 \times 3$ conv, stride 2, pad 1               & $64\times 104\times 104\times 104$                       \\ \hline
$128 \times 3 \times 3 \times 3$ conv, stride 2, pad 1              & $128\times 52\times 52\times 52$                       \\ \hline
 3 consecutive Residual Blocks  $128 \times 3 \times 3 \times 3$ conv & $128\times 52\times 52\times 52$                       \\ \hline
 $64 \times 3 \times 3 \times 3$ conv, stride 2, pad 1 out\_pad 1   & $64\times 104\times 104\times 104$                       \\ \hline
$32 \times 3 \times 3 \times 3$ conv, stride 2, pad 1 out\_pad 1    & $32\times 208\times 208\times 208$                       \\ \hline
$1 \times 7 \times 7 \times 7$ conv, stride 1, pad 3                & $1\times 208\times 208\times 208$                       \\ \hline
\end{tabular}
\end{table}

\begin{table}[htb]
\caption{3D Discriminator structure.}
\centering
\begin{tabular}{|c|c|}
\hline
Layer                                                           & Output feature map size        \\ \hline
Input                                                           & $1\times 208\times 208\times 208$ \\ \hline
$32 \times 4 \times 4 \times 4$ conv, stride 2, pad 1           & $32\times 104\times 104\times 104$                       \\ \hline
$64 \times 4 \times 4 \times 4$ conv, stride 2, pad 1           & $64\times 52\times 52\times 52$                       \\ \hline
$128 \times 4 \times 4 \times 4$ conv, stride 2, pad 1          & $128\times 26\times 26\times 26$                       \\ \hline
$256 \times 4 \times 4 \times 4$ conv, stride 1, pad 1          & $256\times 25\times 25\times 25$                       \\ \hline
$1 \times 4 \times 4 \times 4$ conv, stride 1, pad 1            & $1\times 24\times 24\times 24$                       \\ \hline
\end{tabular}
\end{table}

\begin{table}[htb]\caption{Boundary conditions and physics parameters involved in the GCO process for constructing renal venous trees. Note that $Q_t$ is derived by equal terminal flow assumption:  $Q_t = \frac{Q}{N} \approx  \frac{7ml/min}{68564} \approx 1.7 \times 10^{6} \mu m^3 \, s^{-1} $. Other parameters are identical from [2].}
\centering
\begin{tabular}{|c|c|c|}
\hline
Parameter          & Meaning                                      & Value/Sampling distribution \\ \hline
$r_0$                  & radius of terminal veins & $ \sim \mathcal{N}(10.79, 2.41)$     \\ \hline
$N$                & number of terminal veins             & $ \sim \mathcal{N}(68564, 16647)$      \\ \hline
$Q$                  & inlet flow to renal venous tree              & $7ml/min= 1.167 \times 10^{11} \mu m^3 \, s^{-1}$       \\ \hline
$Q_t$                 & terminal flow out of renal venous tree             & $1.7 \times 10^{6} \mu m^3 \, s^{-1}$       \\ \hline
$\mu$ & viscosity                                    & $3.6 \times 10^{-3} \mathrm{~Pa} \, \mathrm{~s}$      \\ \hline
$w_c$                   & weight factor for material cost      & $\{5, 6, 7\} \times 10^{-8} N \, {\mu m^{-2}} s^{-1}$       \\ \hline

\end{tabular}
\end{table}






\end{document}